# Hierarchical deep reinforcement learning controlled three-dimensional navigation of microrobots in blood vessels

Yuguang Yang,[a,b] Michael A. Bevan,[b] Bo Li[a*]

[a] *Institute of Biomechanics and Medical Engineering, Applied Mechanics Laboratory, Department of Engineering Mechanics, Tsinghua University, Beijing 100084, China*
[b] *Chemical & Biomolecular Engineering, Johns Hopkins University, Baltimore, MD 21218*

**Abstract**

Designing intelligent microrobots that can autonomously navigate and perform instructed routines in blood vessels, a complex and crowded environment with obstacles including dense cells, different flow patterns and diverse vascular geometries, can offer enormous possibilities in biomedical applications. Here we report a hierarchical control scheme that enables a microrobot to efficiently navigate and execute customizable routines in blood vessels. The control scheme consists of two highly decoupled components: a high-level controller setting short-ranged dynamic targets to guide the microrobot to follow a preset path and a low-level deep reinforcement learning (DRL) controller responsible for maneuvering microrobots towards these dynamic guiding targets. The proposed DRL controller utilizes three-dimensional (3D) convolutional neural networks and is capable of learning control policy directly from a coarse raw 3D sensory input. In blood vessels with rich configurations of red blood cells and vessel geometry, the control scheme enables efficient navigation and faithful execution of instructed routines. The control scheme is also robust to adversarial perturbations including blood flows. This study provides a proof-of-principle for designing data-driven control systems for autonomous navigation in vascular networks; it illustrates the great potential of artificial intelligence for broad biomedical applications such as target drug delivery, blood clots clear, precision surgery, disease diagnosis, and more.

**Keyword**: microrobot | artificial intelligence | autonomous navigation | deep reinforcement learning

---

[*] Corresponding author.
 E-mail address: libome@tsinghua.edu.cn (B. Li)

Deploying tiny devices into human body to hard-to-reach locations has been a constant endeavor of mankind to fight diseases. Modern minimally invasive surgeries, in which catheters are navigating in body ducts and vessels, have demonstrated significant advantages over traditional open surgeries in reducing the trauma and infection and speeding up recovery. Along this line, there has been surging interest since last decade in engineering a wide variety of micro-/nano-scale robots for applications in healthcare and biomedicine (1, 2). Researchers are actively exploring possibilities (3, 4) to engineer novel micro/nano-robots that have customizable dynamics (5), bio-compatible and multi-functional surface properties (6), and ability to exploit different power sources (7) such that they work in complex environments for emerging applications like drug delivery and precision surgery. These robots hold the promise of bringing to reality the bold vision described in the 1966 film Fantastic Voyage—shrinking a submarine to microscopic size to venture into blood vessels to cure diseases.

Autonomous microrobots can be used in the blood vessels to carry out specialized tasks including fighting against targeted biological threats, clearing blood clots, and diagnosing diseases from important blood-related biomarkers. For example, the deployment of autonomous microrobots to capture cancer cells in the blood and lymphatic systems can provide a new route to suppress fatal cancer metastasis (8). While macro-scale intelligent robots execute programmable, customizable routines have been well-commercialized (e.g., vacuum cleaning robots, driverless package delivery robots, crime-fighting patrol robot), the autonomous navigation of a microrobot in vascular structures or even *in vitro* remains a daunting task. The major component in the blood is red blood cells (RBCs) that have volume fractions ranging from 8% to 50% in the vessels. The concentrated RBCs acting as obstacles and traps present a considerable hurdle for microrobots to reliably execute instructed tasks in the blood. Navigation of robots in the vascular system *in vivo* is further complicated by the limited visibility, the fast and complex unsteady flows, the unpredictable, rich dynamics of RBCs (e.g., aggregation and deformation), and diverse vessel network topology and geometry. Additionally, navigation in the lymph vessels requires steering through dynamical trapping valve structures not present in blood vessels. In these *bona fide* scenarios, designing a successful autonomous microrobot that can faithfully execute instructed routines must overcome pronounced challenges arising from the unknown and unsteady environments, large-scale navigation distance, diverse obstacles and traps, as well as additional requirement on timing and navigation path.

Here we present a proof-of-principle study on constructing a hierarchical control scheme, which consists of a high-level controller to set dynamic sub-goals to globally guide the microrobot and a low-level deep reinforcement learning (DRL) controller to maneuver robots locally, as a potential solution to the three-dimensional (3D) efficient, programmable navigation of self-propelled microrobots in blood vessels. We show that using raw 3D sensor data, instead of human handcrafted features, the low-level DRL controller can harness deep neural networks to learn robust, efficient and generalizable navigation strategies within diverse blood vessel environments. The hierarchical control design also mimics the hierarchical design making of human beings and offers great flexibility to customize navigation routines in large-scale, complex environments; it ultimately addresses the navigation challenges arising from a broad range of biomedical applications, such as target drug delivery, blood clots clear, precision surgery, disease diagnosis in blood vessels.

**Model and Algorithm**

We establish a hierarchical control scheme to address the 3D navigation of self-propelled microrobots in blood vessels (Fig. 1A). Our hierarchical control scheme consists of a high-level controller dynamically setting short-ranged targets along a desired path (length scale >100 um) (Fig. 1B) and a low-level DRL controller responsible for navigating robots to circumvent RBC obstacles (length scale < 10 um) and moving towards the specified dynamic targets using local observation (Fig. 1C and D). The choice of the DRL controller is motivated by its remarkable performance in sequential decision-making arising from various challenging situations such as games (9, 10) and robotics (11), as well as our recent success in applying DRL to learn generalizable navigation strategies in microstructured environments (12, 13).

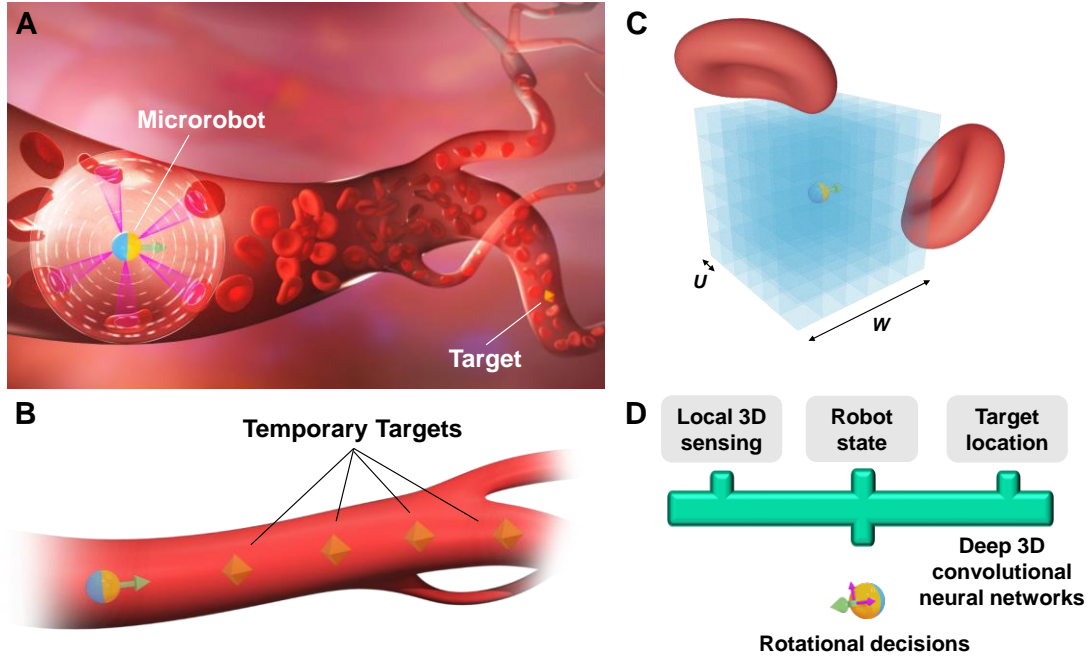

**Figure 1. Hierarchical control scheme for autonomous microrobot navigation.** (A Schematic representation (not to scale) of the low-level controller steering a microrobot to navigate in a blood vessel. Our DRL algorithm employs deep neural networks to take 3D sensing of the microrobot's neighborhood, microrobot's state (position and orientation), and target (octahedron) location as inputs and output rotational decisions. The details of the architecture are provided in the *Methods*. (B) Scheme of 3D local sensation around the microrobot. The sensation is represented by a 3D binary image with width *W* and resolution (pixel size) *U*. The 3D binary image takes value 1 if the central point of the pixel is in a RBC or is outside the vessel, and 0 otherwise. (C) A target generator as a high-level controller to sequentially generate short-ranged targets (octahedrons) that guide the microrobot along a prescribed path. (D) Local 3D sensory input, microrobot state (position and orientation) and target position are fed into a neural network, which outputs the rotational decisions to steer the microrobot towards the target.

Given a preset path in 3D space, the high-level controller selects a point in the path near the microrobot as the *temporary* target position for the low-level controller (Fig. 1B). Once the microrobot get closer to the temporary target, a new farther target alone the path is selected. As the robot continues to follow these guiding targets, it will approximately follow the designed path. Mathematically, let the path be represented by a parametric function $\mathbf{T}(q) \in \mathrm{R}^3$, the sequentially generated targets are then given by $\mathbf{T}(q_1)$, $\mathbf{T}(q_2)$, ..., $\mathbf{T}(q_N)$, where $q_1 < q_2 < \cdots < q_N$, $\mathbf{T}(q_N)$ denote the final target point, or the endpoint the desired path. The generation of new temporary targets is well paced with the progress that the microrobot makes towards these temporary targets, as summarized by Algorithm 1 in *Methods*.

Given a short-ranged target specified by the high-level controller (Fig. 1B), the low-level DRL controller aims to steer the microrobot to the specified target within the minimum time. In this work, we consider a type of microrobot that is engaged in constant self-propulsion but allows continuous control of orientation via different stimuli (e.g., electric (14) and magnetic fields (3, 15), asymmetric shapes (5, 16), flexible structure mechanics (17)). More formally, we denote $v_{SP}$ as the constant propulsion speed and $\mathbf{w}=(w_1, w_2)$, $-w_{max} < w_1, w_2 < w_{max}$, as the two control inputs that change the self-propulsion direction $\boldsymbol{p}$ on two orthogonal basis. Here we define the system state $s$ to consist of microrobot's state (position $\mathbf{r}$ and orientation $\boldsymbol{p}$), the target position $\mathbf{r}^t$ and denote its local 3D observation by $\phi(s)$ (the 3D binary image of the microrobot's neighborhood with a range of approximate 15 μm, double size of a typical RBC).

To seek an optimal control policy $\pi$ that map the system state $s$ and local observation $\phi(s)$ to rotational decisions $\mathbf{w}$, we maximize the expected reward collected during a navigation process $\mathbb{E}\sum_{n=0}^{\infty} \gamma^n \left[ R(s_{n+1}) \right]$ in the policy space (18, 19), where $R$ is the instant reward function encourages or penalizes the system states, $\gamma$ is the discount factor, and $n$ denotes the time step. In the DRL framework, we define the optimal $Q$ function associated with the reward collecting process as

$$Q^*(\phi(s),\mathbf{w}) = \mathbb{E}\left[ R(s_1) + \gamma^1 R(s_2) + \gamma^2 R(s_3) + \cdots \mid \phi(s_0) = \phi(s), \mathbf{w}_0 = \mathbf{w}, \pi^* \right], (1)$$

which is the expected sum of rewards along the navigation process by following the optimal policy $\pi^*$, after observing $\phi(s)$ and making a rotational decision of $\mathbf{w}$. Given $Q^*$ function, the optimal policy is connected to optimal $Q$ function via $\pi^* = \mathrm{argmax}_v$

$Q^*(\phi(s), v)$.

The navigation policy is optimized through the deep deterministic gradient descent algorithm (20), which simultaneously trains a deep neural network, called Critic network, to approximate the optimal $Q$ function, and another deep neural network, called Actor network, to approximate the policy $\pi^*$ (*SI Appendix*, Fig. S1). Both neural networks employ 3D convolution neural layers to process 3D local sensory input and a fully connected layer to process the system state. We train the neural network extensively to estimate $Q^*$ through multiple episodes of navigation in different blood environments (see *SI Appendix* for details; Figs. S2 and S3) to learn robust and generalizable navigation strategies in various scenarios (different RBC configurations, vessel sizes, and target locations).

## Results and Discussion

### Free space navigation

We first examine the free space navigation strategies learned by the DRL controller. Fig. 2 shows the rotational speeds (normalized by the maximum allowed rotation $w_{max}$) parameterized by target locations. For clarity, we place the microrobot at the origin and align its self-propulsion direction with the *x* axis. The *in-plane* rotation changes the self-propulsion direction in the *xy* plane while *out-of-plane* rotation changes the self-propulsion direction in the *xz* plane. Analogous to steering a vehicle, in order to reach the target, the microrobot constantly adjusts its propulsion direction according to the relative position of target throughout the navigation process. Considering targets are in the *xy* plane, key aspects of the navigation strategy are summarized as follows (Fig. 2A and B): (i) When the target is in the front, propulsion direction adjustment is achieved mainly through in-plane rotation in proportion to the angle deviation; (ii) If the target locates behind the microrobot, both in-plane and out-of-plane rotation are engaged at nearly the maximum value in order to quickly reorient the propulsion direction.

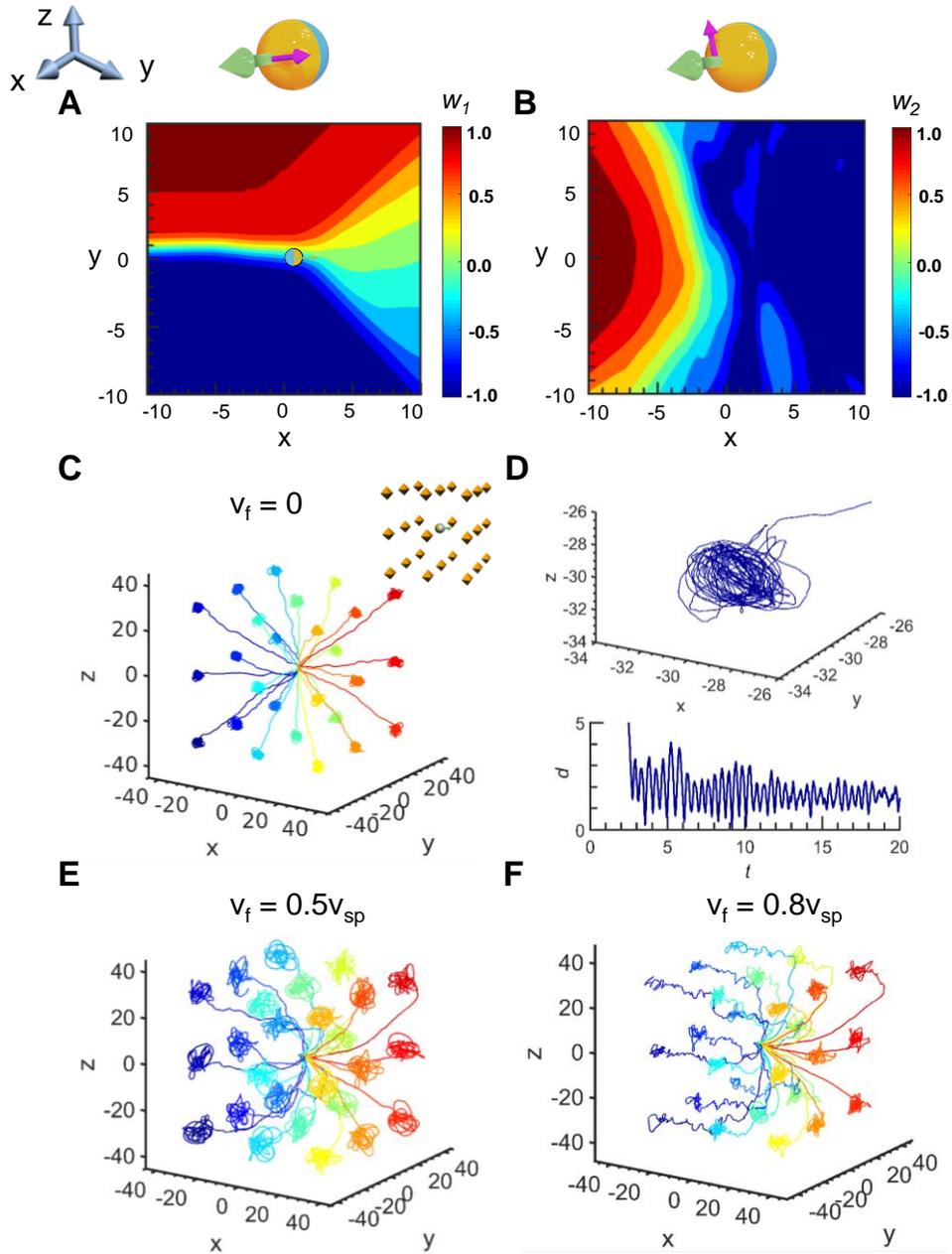

**Figure 2.** Navigation in free space. (A) Learned rotational decision on in-plane rotation speed $w_1$. (B) Learned rotational decision on out-of-plane rotation speed $w_2$. $w_1$ and $w_2$ are normalized by the maximum rotation speed $w_{\max}$. In presenting the control policies, we fix the microrobot at the origin with orientation pointing along the +x direction. The target locations vary in the $xy$ plane. (C) Representative controlled trajectories (200 control steps or 20 $\tau$) of the microrobot (initially located at the origin) navigating towards different target locations. The targets are arranged like lattice, with locations of (-30, -30, -30), (-30, -30, 0), (-30, -30, 30), …, (30, 30, 30). (D) A representative localization trajectory of the microrobot around the target located at (-30, -30, -30) (upper panel); the distance vs. time between the microrobot and the target is showed at the lower panel. (E, F) Representative microrobot trajectories navigating towards different target locations with external flow speeds $v_f = 0.5\ v_{sp}$ (E) and $v_f = 0.8\ v_{sp}$ (F). The setup is the same as (D) except for an existing external flow in the x direction.

The resulting controlled trajectories of the microrobot navigating to targets at different locations are shown in Fig. 2C and Movie S1, where we arrange targets like a lattice cage surrounding the microrobot for comprehensive testing (see the 3D scheme). For targets lying in front (i.e., $\mathbf{r}^t = (30, 0, 0)$), the microrobot directly navigates towards the target. For other target locations that the microrobot does not initially point to, rotational actions are first engaged to quickly reorient the microrobot towards the target and thereafter used to maintain the direction against Brownian motion. In either situation, nearly straight-line trajectories are produced, suggesting the optimality of the navigation strategy (21). Despite the presence of Brownian motion, rotational actions are engaged to correct the disturbance to maintain its trajectory moving towards the target.

The learned control policy enables not just swift navigation towards the target but also the stable localization around the targets after arrival (Fig. 2D). Because the propulsion is constantly engaged, after arriving at the target, the microrobot still needs to constantly adjust its orientation in order to get back to its target. As the microrobot carefully hovers around the target, the microrobot orbits periodically to trace out circular trajectory patterns (radius ~ $v_{SP}/w_{max}$).

So far we have demonstrated the learned control policy under one hyperparameter setting (i.e., $v_{SP}$, $w_{max}$). Control policies under other hyperparameter settings can be obtained via simple arithmetic transformation (*SI Appendix*, Eq. (S3) and Fig. S4] without retraining the model. Moreover, the control policy under external fluid fields can be derived accordingly by treating the system as if a microrobot navigating in still flow field but with a change in its hyperparameter (*SI Appendix*, Eq. (S4)). We applied a flow field in the *x* direction and verified the derived control policies (Fig. 2E and F, *SI Appendix*, Fig. S5). Despite the adversarial impact of external fluid flow, the microrobot can all eventually reach prescribed targets located at different locations. The external fluid flow asymmetrically affects the microrobot motion: it speeds up the microrobot when the microrobot travels along the flow direction but slows down the microrobot if the microrobot travels against the flow direction. Therefore, the optimized navigation trajectories no longer resemble straight lines but are bent towards the flow direction, which is also predicted by theoretical optimal trajectory of micro-swimmers in simple flows (21) Particularly when the magnitude of the flow increases to $v_f = 0.8 v_{SP}$, the trajectories are strongly bent as the microrobot is struggling towards the target. The presence of flow fields also causes delayed arrivals when microrobots travel against the flow (*SI Appendix*, Fig. S5) as well as additional disturbance to the localization process. The radii of the hovering trajectories are significantly larger than the ones when fluid flow is absent. Note that when $v_f$ is greater than the propulsion speed $v_{sp}$, the microrobot is no longer controllable.

**Navigation in blood vessels**

Navigating in blood vessels meets additional challenges as biconcave RBCs and vessel walls can act as traps and barriers. As a first step to evaluate the learned navigation strategy, we consider steering microrobots in a simple environment with a few RBCs in a vessel (Fig. 3A and B). We arrange targets at different locations as in free navigation test in Fig. 2C and examine if the steered microrobot can circumvent RBC obstacles in the way. As shown in the representative trajectories in Fig. 3A, when there is no RBC blockage in the way, the microrobot follows nearly the ideal straight-line path to the target as in the free-space navigation; On the other hand, when an RBC is blocking the direct path, the microrobot will re-orient to get around the RBC. After the arrival, the microrobot employs similar localization strategies around the target as in the free space navigation. To investigate the impact of the vessel wall confinement on navigation, we performed a similar evaluation near the vessel. As shown in Fig. 3B, the microrobot successfully arrived at all the targets near a curved vessel wall. Particularly, when a near-wall RBC is blocking the path to the target, the microrobot will re-orient to circumvent the RBC and simultaneously avoid colliding with the vessel wall.

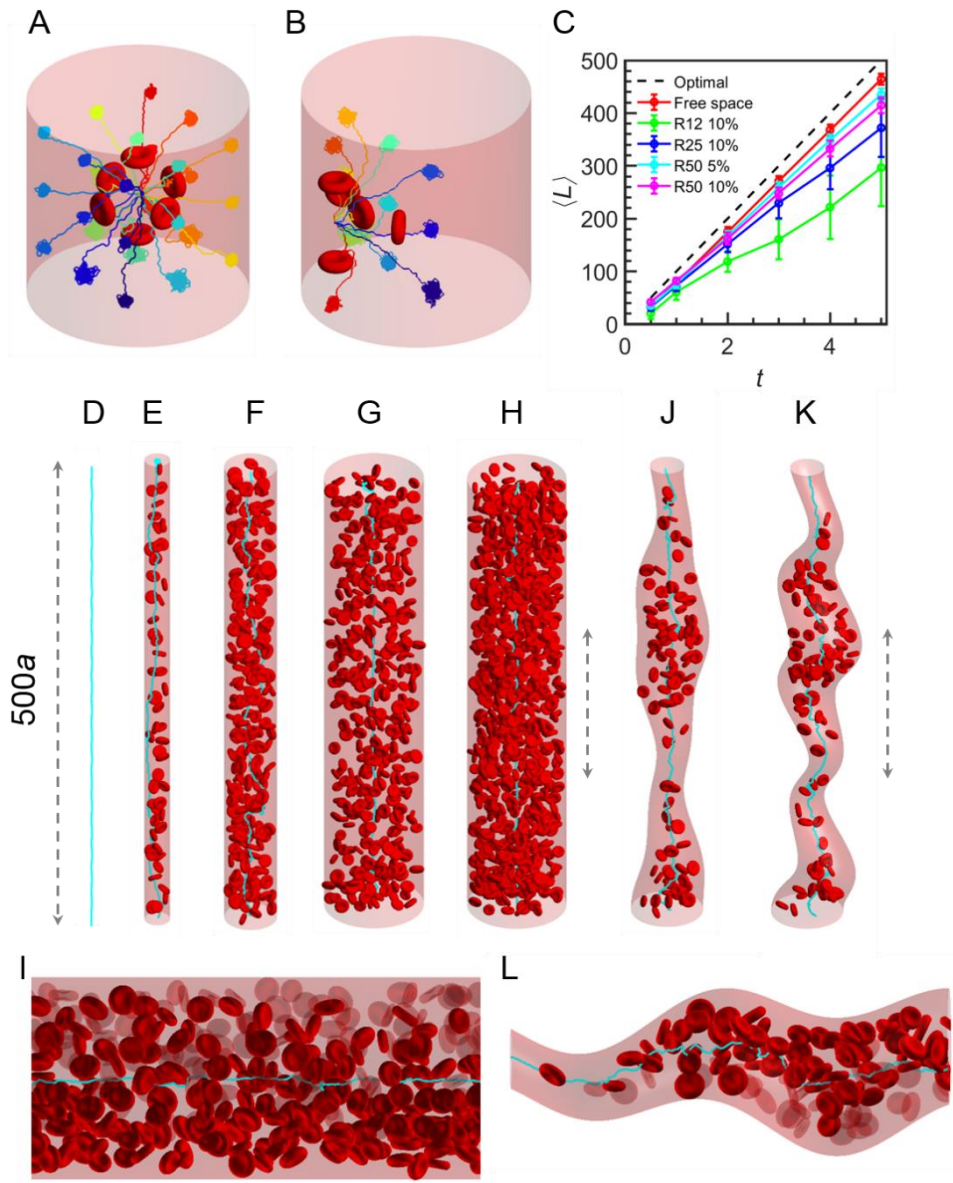

**Figure 3.** Navigation trajectories of a controlled microrobot in blood vessels. (A, B) Representative controlled trajectories of the microrobot navigating towards different targets when the microrobot is initially placed at the center of the vessel (A) or near the boundary of the vessel (B) with orientation point along the +*x* direction. The vessel has a diameter of 50 μm. (C–I) Testing results microrobot navigation performance in blood vessels with different vessel diameters *D* and RBC volume fractions *f*. Microrobots are navigating from the bottom to the top, spanning 500*a* (200 μm). Specifically, (D) free space baseline. (E) *D* = 12 μm, *f* ~ 10%; (F) *D* = 25 μm, *f*~10%; (G) *D* = 50 μm, *f*~5%; (H) *D* = 50 μm, *f*~10%. (I) shows the magnified view of the trajectory in (H). Mean traveled distances vs. time at navigation scenarios (D–H) are collected in (C). (J, K) Example navigation of microrobot in curved blood vessels with varying cross-section diameter. (L) is the magnified view of the trajectory in (K). Here, the RBCs have the diameters uniformly sampled between 6 μm to 8 μm, and the microrobot has diameter 2*a*=1 μm.

Now we evaluate the robustness, generalization, and efficiency of navigation strategies in more realistic blood environments in Fig. 3C–K, which have typical sizes of arteries or veins and different volume fractions of RBCs The major assumption for these blood model environments is that the blood flow is spatially uniform rather than turbulent, such that all objects in the blood flow have similar drifting speeds and appear to be still relative to each other. Note that high viscosity and small diffusivities (due to large sizes of RBCs) will make RBCs appear effectively static compared to the fast-moving self-propelled robots.

We randomly placed RBCs with different configurations (position and orientation) and sizes (uniformly sampled between diameter 6 μm to 8 μm) to create these unseen blood environments to test the generalization of learned strategies. The high-level controller sequentially generates temporary targets to guide the microrobot to follow a straight path extending from the bottom to the top and aligned with the axis of vessels (Algorithm 1). The vast majority of robots can navigate through the vessels by circumventing all RBCs in the way (Movies S2 and 3). Since RBC configurations are randomly generated and are *unseen* in the training stage of the neural network, this test suggests that the neural network learned a generalizable navigation strategy.

We further quantify the navigation performance in blood vessels by calculating the mean travel distance $\langle L \rangle$ versus the mean time $t$ when we set the target at the end of the vessel in Fig. 3C. As a benchmark, in a deterministic limit, the theoretical optimal performance is given by $\langle L \rangle = v_{SP}t$. A rough linear relationship indicates that the microrobot can navigate through a different portion of the vessel with a similar speed while the configuration of RBC varies. Particularly, in the limit where there is no RBC in the vessel, the navigation speed achieves the optimal speed. In general, microrobots transport faster in vessels with a larger radius and fewer RBCs. When the sizes of vessels are the same, more RBCs leads to frequent adjustment of orientation and therefore slows down navigation. At similar RBCs concentrations, more confinement resulting from small vessels sizes produces additional difficulties for microrobots to get around blocking RBCs and therefore causes slower navigation.

As a further test of robustness and generalization of learned navigation strategies, we also examine the navigation in curved vessels with varying diameters (Fig. 3J–L, Movie S3) from bottom to top. Surprisingly, while microrobots are only trained in cylindrical blood vessels, the generalization of DRL controller enables successful navigation in curved vessels. Here we note that while we have achieved decent performance across different blood environment using a single neural network, additional performance gain can be expected if the neural network is further fine-tuned to a specific blood environment, which is the topic of future studies.

Above results assume that the RBCs and the microrobots are experiencing the same ambient flows and therefore the RBCs appears still respect to the microrobot. An extra robustness test is to allow the microrobots to experience an additional external flow field with speed $v_f$. We found that microrobots are capable of arriving at targets when the external flow speed $v_f$ is small ($v_f \leq 0.5\ v_{sp}$) and RBCs are dilute (e.g., 5%) via a simple control policy remapping (*SI Appendix*, Fig. S6, Eqs. (S3) and (S4))

**Exhaustive patrol in blood vessels**

We have demonstrated that the present hierarchical DRL controller can steer the microrobot towards specified targets in both RBC-absent and RBC-present environments. To further demonstrate the strength of our hierarchical control scheme that allows controlled navigation according to a preset routine, we consider the problem of steering a microrobot to exhaustively patrol a blood vessel, analogous to a vacuum robot cleaning a room. The capacity of fast and thoroughly exploring a blood vessel is crucial for applications such as deploying robots to patrol certain regions, to search and clean sparse hidden biological threats (e.g., cancer cells, toxin, etc.) or to fast release and mix drug in complex environments.

Here we consider steering the microrobot to closely follow a predefined path $\mathbf{T}(s)$ = $(x(q), y(q), z(q))$, given by a parametric function

$$\begin{cases} x = R_0 \cos(k_2 q) \cos(k_3 q), \\ y = R_0 \cos(k_2 q) \sin(k_3 q), \\ z = k_1 q, \end{cases} \quad (2)$$

where $q \geq 0$, $k_2$ and $k_3$ determine the projection pattern of path onto the $xy$ plane, $R_0$ denotes the coverage range, $k_1$ determines how fast the path elevates in the $z$ direction. We choose $k_1 = 5$, $k_2 = 5$, $k_3 = 7$, $R_0 = 45$, and the 3D trajectory and its projection on the $xy$ plane is showed in Fig. 4A (Movie S4). By gradually increasing parameter $q$, the curve $(x(q), y(q), z(q))$ traces out a multi-helix pattern elevating from one end to the other end that aims to guide the microrobot to sufficiently sampling the space in a vessel (Fig. 4A). As a baseline case, in a vessel without RBCs, the controlled robots can follow the predefined path with high fidelity, with random deviation quickly corrected by the control policy (Fig. 4B). In vessels with RBCs, the microrobot can manage to closely adhere to the prescribed path by circumventing RBCs in the way (Fig. 4C and D). As RBCs get denser (e.g., 10%), the microrobot needs to deviate from the ideal prescribed path more frequently and the trajectories (top view, Fig. 4A-D) appear to be chaotic

Since microrobots are performing patrol routines from the bottom to the top, the efficiency of routine execution can be measured by elevation speed in $z$ axis (Fig. 4E). The theoretical optimal elevation speed is obtained by assuming a deterministic microrobot with speed $v_{sp}$ exactly follows the preset path (Eq. (2)). In all blood environments, we observe a rough linearity in elevation versus time, indicating microrobots are making constant progress in the patrol task. With 0%, 5%, and 10% RBCs in the vessels, the elevation speeds are 85.9%, 73.9%, and 64.2%, respectively, of the optimal speed, as a result of slowdowns caused by more RBC blockages.

The ultimate quantification of the routine execution quality is measured by distance between the 3D preset path $\mathbf{T}$ (Eq. (2)) and the actual executed path $\mathbf{r}$ after appropriate alignment. Particularly, we can define point-wise deviation between the two paths at an arbitrary point $q$ by

$$\Delta(q) = \|\mathbf{r}(q) - \mathbf{T}(q')\| \quad (3)$$

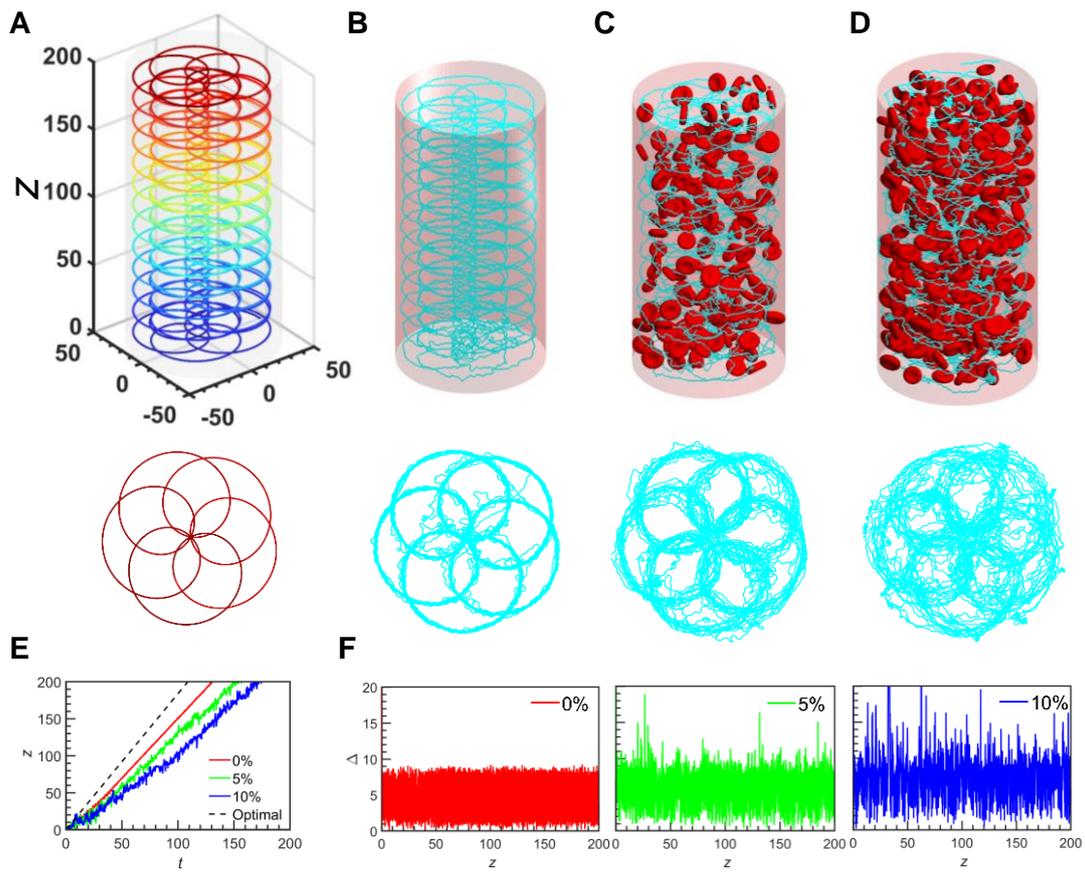

**Figure 4.** Exhaustive patrol in a blood vessel. (A) The predefined path that aims to exhaustively patrol a vessel. (B–D) Controlled trajectories following the preset path to exhaustively patrol the vessels. (B) zero RBCs, (C) ~5% RBCs, (D) ~10% RBCs. Upper panel: 3D view; Lower panel: Top view. (E) Elevation z vs. time of patrolling trajectories in different blood environments. (F) Pointwise path distance Δ vs. elevation z in different blood environments (Eq. (3)).

where *q'* is the corresponding optimal alignment in *T* computed using the dynamic time warping algorithm[†]. The mean deviation between the two paths is given by averaging enough sample points within the paths (*SI Appendix*). As shown in Fig. 4F, the pointwise deviation Δ at different elevations for all patrolling cases are fluctuating around increasing mean deviation of 4.8*a*, 6.0*a*, and 7.2*a* in blood environments with 0%, 5%, and 10% RBCs. With increasing RBCs, occasional spikes in Δ are more frequent since microrobots are going extra way to get around. Overall, our control scheme can enable microrobots to execute patrol routines within different microstructured environments with high fidelity. Moreover, by modifying the patrol path defined by Eq. (2), different patrol strategies such as adaptive exploration in vessels with varying size can be implemented.

---

[†] Dynamic time warping could be directly computed via MATLAB https://www.mathworks.com/help/signal/ref/dtw.html

**Model analysis**

We now decode what has been learned in the decision-making module enabled by DRL to understand the navigation performance in above tasks. In a toy blood environment (Fig. 5A), we apply *t*-Distributed Stochastic Neighbor Embedding (*t*-SNE) algorithm (22) to embed the learned representations of randomly sampled states into a 2D plane and color each point by the state value given by.

$$V(s) = \max_{\mathbf{w}} Q^*(\phi(s), \mathbf{w}) = \mathbb{E}\left[\sum_{n=0}^{\infty} \gamma^n R(s_{n+1}) \mid s_0 = s, \pi^*\right]. \tag{4}$$

The state value provides information on if one state *s* is favorable to another state; a higher *V* indicates the controlled microrobot can arrive sooner than states with lower *V*.

We consider five configurations ①–⑤ in Fig. 5 to examine how the network perceives different situations. As shown in Fig. 5A, high dimensional system states are embedded in the 2D plane apparently based on the shortest path distance to the target location, with closer states on the right. For example, configuration ① with the closest distance to the target has its embedding on the right. Similarly, in configurations ② and ④, where the microrobot in ② gets blocked by the RBC and has to re-orient to get to the target and the microrobot in ④ without RBC blockage in the way, the two configurations got assigned similar value. Additionally, in configurations ③ and ⑤, a microrobot in ⑤ gets blocked by two RBCs in the way but the configuration is evaluated to have similar state value to ⑤ where the microrobot is not blocked by any RBCs. We hypothesize that the neural network can implicitly estimate the shortest paths based on local sensor information and the target position, and uses this estimate to guide the rotation decision to follow these shortest paths. To validate this hypothesis, we use Dijkstra to estimate the shortest path distance from each state to the target. Under this hypothesis, the shortest path distance can provide state value estimation in Eq.(4) via

$$V^{est}(s) = \gamma^{l_S/v_{SP}t_c}, \tag{5}$$

where $\gamma$ is the discount factor used in Eq. (4), $l_S$ is the shortest path length from the microrobot's position to the specified target, and $l_S/v_{SP}t_c$ is the number of control steps needed to move along the shortest path. The similarity in the learned state value function and the estimated one (Fig. 5B) suggests the microrobot has acquired nearly optimal navigation strategies; this is, making rotation decisions to follow approximate shortest paths. Although we never explicitly model this knowledge, the orientation rotating to follow the shortest path emerges after reinforced learning of extensive navigation data.

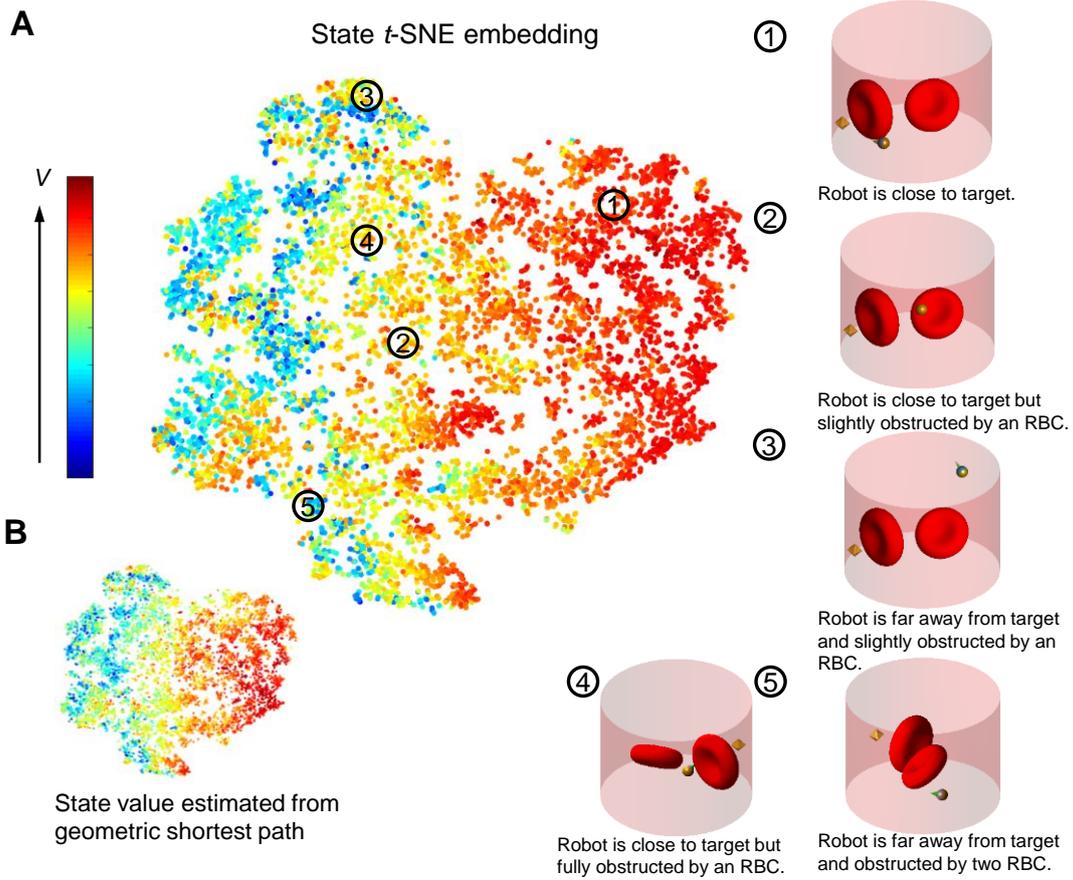

**Figure 5.** (A) The two-dimensional t-SNE embedding of last hidden layer representation of the neutral network in an example navigation task. Every point corresponds to a 2D representation of the internal state associated with the observations at the microrobot states (r, p). Points are colored by the state value. (B) Estimated state value based on the shortest-path estimation (Eq. (5)).

## Conclusion and outlook

We have presented a proof-of-principle study on designing a hierarchical control scheme to enable autonomous, customizable navigation of microrobots in blood vessels. Although the low-level control DRL alone is sufficient for relatively simple navigation scenarios (12), a high-level controller gives more control on the travel trajectories and is suited for designing more flexible automated navigation routines. While we do not attempt at a fully modeling of the real complexity of blood environment, we aim to emphasize the key idea of local sensing, hierarchical planning, and data-driven learning.

We show that a 3D sensor of local environment together with DRL based control can learn robust navigation strategy in complex blood environments. In a series of navigation tasks of different objectives, we show that the controlled microrobot can efficiently travel in blood vessels with different RBC concentrations and configurations and complex vascular geometries and flows. We further demonstrate that the neural network can learn effective representations of observations that underpinning successful navigation performance. Our results not only demonstrate a general data-driven control scheme to enable navigation autonomy in human blood vessels, but also lay the foundation for devising more sophisticated autonomy of nano/microrobot navigation in an ample spectrum of complex environments, either *in vivo* or *in vitro*.

The data-driven nature of deep neural network method enables our control framework for multiple extensions. The proposed algorithm can be combined with the high-fidelity physical simulator of the true blood to learn control strategies that address complexity arising from non-steady blood flow, RBC deformation (23, 24), and hemodynamics (25). Similarly, our control scheme also applies to a broad class of nano-robots in other navigation scenarios like the urinary tracts or the eyeballs in the human body or 3D porous media for environmental applications. The high decoupling nature of our control scheme also allows substitution of low-level control module suited for specific robots and motors. A further extension will be multi-agent stochastic control (26, 27) of multiple microrobots to achieve capable swarm intelligence, such as capturing circulating tumor cells in the blood (28, 29). The sparse and simple sensor design considered herein can indeed simplify the extension to the multi-agent framework.

## Materials and Methods

### Dynamics of microrobot

We model the direction-controllable microrobot by

$$\begin{aligned}\partial_t \mathbf{r} &= \boldsymbol{\xi}_\mathbf{r}(t) + v_{\text{SP}}\mathbf{p}, \\ \partial_t \mathbf{p} &= [\mathbf{I} - \mathbf{p}(t)\mathbf{p}(t)] \cdot \boldsymbol{\xi}_\mathbf{p}(t) + w_1 \mathbf{q}_1 + w_2 \mathbf{q}_2,\end{aligned} \quad (6)$$

where **r** and **p** denote the position and the orientation (which is also the self-propulsion direction), respectively; $t$ is time, and $v_{SP}$ is propulsion speed taking constant value; $\mathbf{w}=(w_1, w_2)$, $-w_{max} < w_1, w_2 < w_{max}$ are the two control inputs that change the self-propulsion direction in two orthogonal basis directions of $\mathbf{q}_1$ and $\mathbf{q}_2$, where $\mathbf{q}_1 = \mathbf{e}_z \times \mathbf{p}$ ($\mathbf{e}_z$ is the unit vector in the z-direction) and $\mathbf{q}_2 = \mathbf{p} \times \mathbf{q}_1$. Brownian translation and rotation are characterized by zero-mean independent multivariate Gaussian noise process $\boldsymbol{\xi}_\mathbf{r}$ and $\boldsymbol{\xi}_\mathbf{p}$ with covariance $E[\boldsymbol{\xi}_\mathbf{r}(t)\boldsymbol{\xi}_\mathbf{r}^T(t')] = 2D_t \mathbf{I} \delta(t-t')$, $E[\boldsymbol{\xi}_\mathbf{p}(t)\boldsymbol{\xi}_\mathbf{p}^T(t')] = 2D_r \mathbf{I} \delta(t-t')$, where $D_t$ is the translational diffusivity, $D_r$ is the rotational diffusivity, and $\mathbf{I}$ denotes the unit tensor. All lengths are normalized by microrobot radius $a$ and time is normalized by $\tau = 1/D_r$. The control update time is $t_c = 0.02\tau$, the integration time step $\Delta t = 0.001\tau$, and $D_t = 1.33 a^2 D_r$.

**Hierarchical control algorithm**

The hierarchical control algorithm performs iterations on two levels. The high-level controller iteratively updates temporary short-ranged targets along the desired path. The low-level DRL controller iteratively updates the rotational decisions at an interval of $t_c$ based on the microrobot state and the local observation. The target update is triggered only when the microrobot is making progress, or getting closer, to the target. The complete algorithm is illustrated in **Algorithm 1**. In selecting the temporary targets, $d_s$ is set at 20.

| **Algorithm 1:** Hierarchical control algorithm for microrobot navigation |
|---|
| Given a desired path represented by a parametric function $\mathbf{T}(q) \in \mathbb{R}^3$. Denote microrobot position by $\mathbf{r}$.<br>**While** True:<br>Select a temporary target on the path, denoted by $\mathbf{r}^t = \mathbf{T}(q^*)$, where $q^* = \text{argmin}_q [\|\mathbf{T}(q) - \mathbf{r}\| > d_s]$ and solved $q^*$ is required to be monotonically increasing.<br>    **While** robot is not getting closer to the target:<br>        Steering the microrobot towards the target rt based on DRL policy.<br>    **End While**<br>**End While** |

**Neural network and training procedure for DRL**

Our neural network (Fig. S1) architecture is adapted from the general Actor-Critic deep neural network for reinforcement learning (20) and utilizes convolutional neural network to extract features from 3D sensor input. A similar architecture has been used in our previous work (13). Below is a high-level description of the architecture and see *SI Appendix* for more details.

*Actor network*

The Actor network takes two inputs and output a two-dimensional rotational decision. The first input is the pixel level binary sensory input of 30×30×30 cubic neighborhood centering on the particle and aligned with its self-propulsion direction **p** (pixel width is *a*; 1 denotes the presence of obstacles or out of the boundary, and 0 *vice versa*). The second input is a six-dimensional vector as the concatenation of the target position in the microrobot's local coordinate frame and the self-propulsion direction **p**. The neighborhood sensory input is fed to two 3D convolutional layers (30-32). The local target coordinate first enters a fully connected layer. Then the output from the target coordinate input and the sensory input will merge and enter a fully connected layer. The final output layer is a fully-connected linear layer with two output associated with the choice of $w_1$ and $w_2$. Tanh nonlinearity is applied to constrain the two output to within [-1, 1].

*Critic network*

The Critic network takes three inputs and output a scalar $Q$ value. The first input is the binary cubic image of the neighborhood fed into two layers of 3D convolutional layers (same as that in the actor network). The local target and the self-propulsion director **p** will first concatenate with the action output (two dimension) from the actor network. The 8 dimensional concatenated vector then will enter a fully connected layer. Then the output from the target coordinate input and the sensory input will merge and enter a fully connected layer of 64 unit following by rectifier nonlinearity. The output layer is a fully-connected linear layer with one output as the $Q$ value given input of observation and action.

*Training algorithm and procedures*

The algorithm we used to the train the neural network is the deep deterministic policy gradient algorithm (20) plus the hindsight experience replay enhancements (12,

33) for data augmentation, and scheduled multi-stage learning to boost sample efficiency (following the idea of curriculum learning (34)). At the beginning of each episode, the initial microrobot state and the target position are randomly generated in a scheduled manner. As training proceeds, the initially distance between microrobot and the target is gradually increased from a smaller value to a cap value. This allows the neural network to gradually acquires control policies of increasing difficulties (in terms of distance to the target). To alleviate the exploration-exploitation dilemma (19), during the training process, we add noises to the actions from actor network to enhance the exploration in the policy space. The noise is sampled from an Ornstein–Uhlenbeck (OU) process (on each dimension). There are two loss functions we used to train the Actor network and the Critic network, respectively. By minimizing the loss function associated with the Critic network, the Critic network is optimized to approximate the optimal $Q$ function; By minimizing the loss function associated with the Actor network, the Actor network optimizes the approximated $\pi$. The complete algorithm is given below. See *SI Appendix* for more details.

## SI Appendix

**Fig. S1**. The Actor-Critic architecture used to learn optimal control policies for the microrobot.
**Fig. S2**. The approximate RBC shape we used to facilitate fast collision check in the simulation.
**Fig. S3.** Scheme of training and evaluation workflow. Efficient training is achieved by selecting increasingly challenging tasks as motivated by curriculum learning.
**Fig. S4.** Navigation and localization trajectories of microrobots in free space with different propulsion rotation ratio.
**Fig. S5.** Navigation and localization trajectories of microrobots in free space with different external flow fields.
**Fig. S6.** Navigation of microrobots in blood vessels with different external flow fields.
**Movie S1.** Navigation and localization of microrobots in free space with different external flow fields.

**Movie S2.** Navigation of microrobots in blood vessels with different vessel diameter $D$ and RBC volume fraction $f$. Specifically, from left to right, $D = 12 um, f \sim 10\%$; $D = 25 um, f \sim 10\%$; $D = 50 um, f \sim 5\%$; $D = 50 um, f \sim 10\%$.

**Movie S3.** Navigation of microrobots in curved blood vessels with varying cross-

section diameter.

**Movie S4.** Exhaustive search in a blood vessel. From left to right, the blood vessel has zero RBCs, ~5% RBCs, and ~10% RBCs.